\documentclass[twocolumn]{revtex4}
\usepackage{epsfig}
\usepackage{amssymb}
\usepackage{amsmath}
\usepackage{amsfonts}
\usepackage{graphicx}
\usepackage{mathrsfs}
\usepackage[dvips]{color}
\usepackage{multirow}

\newcommand{\vp}{{\vec p}}

\newcommand{\vd}{{\vec \partial}}

\newcommand{\R}{\mathbb{R}}
\newcommand{\C}{\mathbb{C}}

\newcommand{\fn}{{\mathfrak{n}}}

\newcommand{\fz}{\mathfrak{z}}

\newcommand{\bbr}{\mathbf{r}}

\newcommand{\bH}{\mathbf{H}}
\newcommand{\bI}{\mathbf{I}}

\newcommand{\bM}{\mathbf{M}}

\newcommand{\bS}{\mathbf{S}}

\newcommand{\bU}{\mathbf{U}}

\newcommand{\cK}{\mathcal{K}}

\newcommand{\cX}{\mathcal{X}}
\newcommand{\cY}{\mathcal{Y}}
\newcommand{\cZ}{\mathcal{Z}}
\newcommand{\be}{\begin{equation}}
\newcommand{\ee}{\end{equation}}
\newcommand{\bea}{\begin{eqnarray}}
\newcommand{\eea}{\end{eqnarray}}
\newcommand{\nn}{\nonumber}

\newcommand{\ed}{\end{document}}

\newcommand{\bi}{\begin{itemize}}
\newcommand{\ei}{\end{itemize}}

\newcommand{\bce}{\begin{center}}
\newcommand{\ece}{\end{center}}

\newcommand{\sD}{\mathscr{D}}

\newcommand{\sT}{\mathscr{T}}

\begin{document}

\title{Transfer Matrix Formulation of Scattering Theory in Two and Three Dimensions}

\author{Farhang Loran$^*$ and Ali~Mostafazadeh$^\dagger$\\[6pt]
$^*$Department of Physics, Isfahan University of Technology, Isfahan, 84156-83111, Iran\\[6pt]
$^\dagger$Departments of Mathematics and Physics, Ko\c{c} University, 34450 Sar{\i}yer,
Istanbul, Turkey}

\begin{abstract}
In one dimension one can dissect a scattering potential $v(x)$ into pieces $v_i(x)$ and use the notion of the transfer matrix to determine the scattering content of $v(x)$ from that of $v_i(x)$. This observation has numerous practical applications in different areas of physics. The problem of finding an analogous procedure in dimensions larger than one has been an important open problem for decades. We give a complete solution for this problem and discuss some of its applications. In particular we derive an exact expression for the scattering amplitude of the delta-function potential in two and three dimensions and a potential describing a slab laser with a surface line defect. We show that the presence of the defect makes the slab begin lasing for arbitrarily small gain coefficients.\\

\noindent {Pacs numbers: 03.65.Nk, 42.25.Bs}\\

\medskip

\end{abstract}

\maketitle

\section{Introduction}

Scattering of waves has been one of the most important subjects of research in physics for over a century. Most students of physics begin their study of this phenomenon in undergraduate courses of quantum mechanics where they learn time-independent scattering theory and its application in solving simple one-dimensional problems. In more advanced courses they are exposed to the $\bS$-matrix formulation of scattering theory which is of basic importance in quantum field theories and particle physics \cite{weinberg}.

In one dimension, there is an alternative to the $\bS$-matrix, called the transfer matrix, that similarly to the $\bS$-matrix stores all the scattering information about the scattering potential. To give the definition of the transfer matrix, we recall that for a scattering potential $v$, every solution of the time-independent Schr\"odinger equation,
    \be
    -\psi''(x)+v(x)\psi(x)=k^2\psi(x),
    \label{sch-eq-1d}
    \ee
has the asymptotic form:
    \[\psi(x)=A_\pm e^{ikx}+B_\pm e^{-ikx}~~{\rm for}~~x\to\pm\infty.\]
The transfer matrix of $v$ is the $2\times 2$ matrix $\bM$ satisfying
	\be
    \left[\begin{array}{c}
	A_+ \\ B_+ \end{array}\right]=\bM\left[\begin{array}{c}
	A_- \\ B_- \end{array}\right].
	\label{M-def}
	\ee
It is easy to show that the reflection and transmission coefficients of $v$ are given by the entries of $\bM$, \cite{razavy,jpa-2009a,sanchez}.

The main advantage of the transfer matrix over the $\bS$-matrix is that it satisfies an extremely useful composition rule which allows for the reduction of the scattering problem for $v$ to that of any set of constituent potentials  $v_1, v_2, \cdots, v_n$ provided that $v=\sum_j v_j$ and the support $I_j$ of $v_j$ (i.e., the smallest interval outside which $v_j$ vanishes) do not intersect except possibly at a single point. If $\bM_j$ is the transfer matrix of $v_j$ and $I_j$ lies to the left of $I_{j+1}$, the transfer matrix of $v$ is given by
    \be
    \bM=\bM_n\bM_{n-1}\cdots\bM_1.
    \label{comp-rule}
    \ee
This relation makes the transfer matrix into an indispensable tool for dealing with a variety of scattering problems \cite{sanchez}. Typical examples are the study of multilayer optical systems \cite{yeh-LPP-1,yeh-LPP-2,yeh-LPP-3,yeh-LPP-4}, acoustic waves \cite{accoustics-1,accoustics-2,accoustics-3}, transport phenomena in mesoscopic systems \cite{MK}, photonic band structure calculations \cite{pendry-92,pendry-94}, optical potential engineering \cite{pra-2014bc-2015b-2,pra-2014bc-2015b-3,pra-2014bc-2015b-4}, etc. These applications are however limited to (effectively) one-dimensional physical systems. This is because the very definition of a multidimensional transfer matrix, that contains all the information about the scattering features of the potential and fulfils the composition property (\ref{comp-rule}), has been lacking. The purpose of the present article is to introduce such a notion of transfer matrix and use it to develop an alternative to standard approach to scattering theory \cite{sakurai,weinberg-qm} that is capable of dealing with a variety of intricate scattering problems.

The basic conceptual framework for the present work is provided by the recent observation that the transfer matrix in one dimension can be expressed as
    \be
    \bM=\bU(\infty,-\infty),
    \label{M=U-1D}
    \ee
where $\bU(x,x_0)$ is the evolution operator for a fictitious non-unitary two-level quantum system with $x$ playing the role of the evolution parameter \cite{ap-2014,pra-2014a}. This was originally motivated by the fact  that the evolution operator of every quantum system has the same composition (group) property as the transfer matrix, i.e., (\ref{comp-rule}). It turns that one can define a two-component state vector $\Psi(x)$ and an $x$-dependent effective matrix Hamiltonian ${\bH}(x)$ of the form
    \bea
    \Psi(x)&:=&\frac{1}{2}\left[\begin{array}{c}
	e^{-ikx}[\psi(x)-ik^{-1} \psi'(x)]\\[3pt]
    e^{ikx}[\psi(x)+ik^{-1} \psi'(x)]\end{array}\right],
    \label{Psi-1D}\\
    {\bH}(\tau)&:=&\frac{v(x)}{2k}\left[\begin{array}{cc}
	1 & e^{-2ikx}\\
	-e^{2ikx}&-1\end{array}\right],
    \label{H-1D}
    \eea
such that the time-independent Schr\"odinger equation~(\ref{sch-eq-1d}) becomes equivalent to the time-dependent Schr\"odinger equation
    \be
    i\partial_x\Psi(x)=\bH(x)\Psi(x),
    \label{time-dep-sch-eq-1D}
    \ee
with $x$ playing the role of ``time,'' \cite{ap-2014}.

Clearly, there is an infinity of choices for $\Psi(x)$ (and the corresponding $\bH(x)$) that allow for expressing (\ref{sch-eq-1d}) in the form (\ref{time-dep-sch-eq-1D}), \cite{FV-1958,jmp-1998}. Note however that among these, (\ref{Psi-1D}) is the only choice that fulfils the asymptotic boundary conditions:
    \be
    \Psi(x)\to\left[\begin{array}{c} A_\pm\\ B_\pm\end{array}\right]~~{\rm for}~~x\to\pm\infty.
    \label{Psi=asym-1D}
    \ee
This relation is the key to the identification of the transfer matrix $\bM$ with $\bU(-\infty,\infty)$, where $\bU(x,x_0)$ is the evolution operator for the Hamiltonian (\ref{H-1D}), i.e., the operator satisfying
    \be
    \Psi(x)=\bU(x,x_0)\Psi(x_0).
    \label{U-def}
    \ee
Equation~(\ref{M=U-1D}) is a direct consequence of (\ref{M-def}), (\ref{Psi=asym-1D}), and (\ref{U-def}) for $x_0\to-\infty$ and $x\to\infty$.

Another important consequence of (\ref{Psi-1D}) is that the corresponding effective Hamiltonian, namely (\ref{H-1D}), vanishes for the values of $x$ where $v(x)=0$. We can use this observation
to give a derivation of the composition rule for transfer matrices. Let us denote by $\bH_j(x)$ and $\bU_j(x,x_0)$ the effective Hamiltonian and the evolution operator for the constituent potentials $v_j$, respectively. Suppose that the support of $v_j$ is given by $I_j=[a_j,a_{j+1}]$ for some real numbers $a_j$ and $a_{j+1}$. Then for all $x\notin I_j$, $\bH_j(x)=0$. This in turn implies that
whenever the interval $(x_0,x)$ does not intersect $I_j$, we have $\bU_j(x,x_0)=\bI$, where $\bI$ is the $2\times 2$ identity matrix. In particular,
    \be
    M_j=\bU_j(\infty,-\infty)=\bU_j(a_{j+1},a_j)=\bU(a_{j+1},a_j),
    \label{M-j=}
    \ee
where the last equality follows from the fact that $v_j(x)$ is the restriction of $v(x)$ to $I_j$. Next, we recall that because $I_j$ lies to the left of $I_{j+1}$, $a_1\leq a_2\leq a_3\leq\cdots \leq a_{n+1}$. This together with (\ref{M=U-1D}), (\ref{U-def}), and (\ref{M-j=}) imply
    \be
    \bU(\infty,\infty)=\bU(a_{n+1},a_n)\bU(a_{n},a_{n-1})\cdots \bU(a_2,a_1).
    \nn
    \ee
In light of (\ref{M=U-1D}) and (\ref{M-j=}), this equation is equivalent to (\ref{comp-rule}).

The idea of expressing the transfer matrix of one-dimensional scattering theory in terms of the evolution operator for a two-level quantum system has revealed an intriguing manifestation of geometric phases in semiclassical scattering \cite{jpa-2014a}. It has also been used to devise a method of computing higher order corrections to semiclassical scattering calculations \cite{jpa-2014b} and a generalization of the composition rule (\ref{comp-rule}) to situations where the constituting potentials $v_j$ have overlapping support \cite{ap-2015}. In this article, we use this idea as a guiding principle for developing a transfer matrix formulation of the scattering theory in two and three dimensions. Since the basic difficulty that has hindered the progress towards solving this problem is already present in two dimensions, we give a more thorough treatment of scattering theory in two-dimensions. In order to focus on the conceptual developments and their concrete applications, we give the technical details of most of the calculations in the appendices.

\section{Transfer matrix in 2D}

Consider the Schr\"odinger equation,
	\be
	\left[-\partial_x^2-\partial_y^2+v(x,y)\right]\psi(x,y)=k^2\psi(x,y),
	\label{sch-eq}
	\ee
where $v$ is a real- or complex-valued scattering potential and $(x,y)$ are cartesian coordinates in the plane $\R^2$. Let us identify the $x$-axis with the scattering axis, use $\mathbf{r}$ to denote the position vector associated with $(x,y)$, and take the Fourier transform of (\ref{sch-eq}) with respect to $y$. This gives
	\be
	\left[-\partial_x^2+v(x,i\partial_p)\right]\tilde\psi(x,p)=\omega(p)^2\tilde\psi(x,p),
	\label{sch-eq-FT}
	\ee
where $\tilde\psi(x,p):=\int_{-\infty}^\infty dy e^{-ipy}\psi(x,y)$ is the Fourier transform of $\psi(x,y)$ over $y$ and $\omega(p):=\sqrt{k^2-p^2}$. We identify $v(x,i\partial_p)$ with the integral operator defined by $v(x,i\partial_p)\phi(p):=\frac{1}{2\pi}\int_{-\infty}^\infty dq\,\tilde v(x,p-q) \phi(q)$, where $\phi$ is a test function.

For the scattering problem, we are interested in the solutions of (\ref{sch-eq-FT}) that tend to a plane wave for $x\to\pm\infty$. These exist provided that $|p|\leq k$ and $|v(x,y)|$ has a sufficiently rapid asymptotic decay rate. We therefore demand that as $r:=\sqrt{x^2+y^2}\to\infty$, $|v(x,y)|$ tends to zero at such a rate that every solution of (\ref{sch-eq}) has the following asymptotic form for $x\to\pm\infty$:
	\be
    \frac{1}{2\pi}
	\int_{-k}^k dp\, e^{ipy}\left[A_\pm(p)e^{i\omega(p)x}+
	B_\pm(p) e^{-i\omega(p)x}\right].
    \nn
	\ee
Here $A_\pm$ and $B_\pm$ are coefficient functions vanishing outside the interval $[-k,k]$.

We identify the transfer matrix of the potential $v$ with the $2\times 2$ matrix operator $\bM$ satisfying (\ref{M-def}). As we show in the sequel, this notion of transfer matrix not only fulfills the composition rule (\ref{comp-rule}), but also possesses complete information about the scattering properties of $v$.

We begin our analysis of the properties of $\bM$ by introducing the following two-dimensional generalization of the two-component state vector (\ref{Psi-1D}) and effective Hamiltonian operator (\ref{H-1D}).
    \begin{align}
    \Psi(x,p)&:=\frac{1}{2}\left[\begin{array}{c}
    e^{-i\omega(p)x}[\tilde\psi(x,p)-i\omega(p)^{-1}\tilde\psi'(x,p)]\\[3pt]
    e^{i\omega(p)x}[\tilde\psi(x,p)+i\omega(p)^{-1}\tilde\psi'(x,p)]\end{array}\right],
    \label{two-comp}
    \\[6pt]
    \boldsymbol{\bH}(x,p)&:=\frac{1}{2\omega(p)} e^{-i\omega(p)x\boldsymbol{\sigma}_3}
	v(x,i\partial_p)\,\boldsymbol{\cK}\,e^{i\omega(p)x\boldsymbol{\sigma}_3},
	\label{sH=}
    \end{align}
where $\tilde\psi'(x,p):=\partial_x\tilde\psi(x,p)$, $\boldsymbol{\sigma}_i$ are the Pauli matrices, and $\boldsymbol{\cK}:=\boldsymbol{\sigma}_3+i\boldsymbol{\sigma}_2$. $\Psi(x,p)$ has the following two important properties:
    \begin{enumerate}
    \item It satisfies $i\partial_x\Psi(x,p)=\boldsymbol{\bH}(x,p)\Psi(x,p)$, which is equivalent to (\ref{sch-eq-FT}). In other words, $\Psi(x,p)=\boldsymbol{\bU}(x,x_0)\Psi(x_0,p)$, where $\boldsymbol{\bU}(x,x_0)$ is the evolution operator for $\boldsymbol{\bH}(x,p)$.
    \item It tends to $\left[\begin{array}{c}A_\pm(p)\\ B_\pm(p)\end{array}\right]$ for
        $x\to\pm\infty$.
    \end{enumerate}
By virtue of these properties and Eq.~(\ref{M-def}), we can express $\bM$ in the form
	\bea
	\bM&=&
	\boldsymbol{\bU}(\infty,-\infty)=\sT\exp\int_{-\infty}^{\infty}\!\!-i\boldsymbol{\bH}(x,p) dx
	\label{M=exp}\\
	&=&\mathbf{I}-i\int_{-\infty}^\infty\!\!\! dx_1 \boldsymbol{\bH}(x_1,p)\nn\\
    &&+(-i)^2\int_{-\infty}^\infty\!\!\! dx_2\int_{-\infty}^{x_2}\!\!\!  dx_1
	\boldsymbol{\bH}(x_2,p)	\boldsymbol{\bH}(x_1,p)+\cdots,\nn
	\eea
where $\sT$ is the ``time-ordering'' operation \cite{weinberg-qm} with $x$ playing the role of ``time.''

\begin{figure}
	\begin{center}
	\includegraphics[scale=.45]{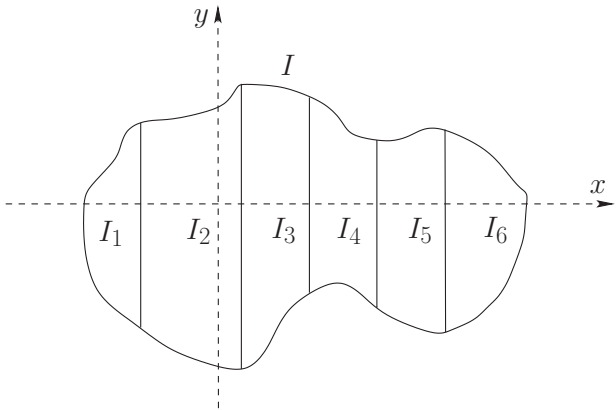}
	\caption{Support $I$ of a finite-range potential (the enclosed region) sliced along the $x$-axis.}
	\label{fig0}
	\end{center}
	\end{figure}

According to (\ref{sH=}), for each $x$ satisfying $v(x,y)=0$, we have $\boldsymbol{\bH}(x,p)=0$. Similarly to the case of one dimension, we can employ this property of $\boldsymbol{\bH}(x,p)$ together with Eq.~(\ref{M=exp}) to establish the composition rule (\ref{comp-rule}). This rule holds true if $I_j$ do not have overlapping projections onto the scattering ($x$-) axis except possibly in a discrete set of points. For example, we can slice the support of $v$ along the $x$-axis, label the slices by $I_j$ (with $I_j$ to the left of $I_{j+1}$ as depicted in Fig.~\ref{fig0}), and take
    \[v_j(\bbr)=\left\{\begin{array}{cc}
    v(\bbr)&{\rm for}~\bbr\in I_j,\\
    0 &{\rm for}~\bbr\notin I_j,\end{array}\right.\]
where $\bbr:=(x,y)$.

In order to see how $\bM$ relates to the scattering amplitude of the potential $v$, we consider a left-incident wave and write the corresponding scattered wave in the form
    \be
    \psi_{\rm scat}(\mathbf{r})=\Theta(-x)\psi_{-}(\mathbf{r})+\Theta(x)\psi_{+}(\mathbf{r})
    ~~~~{\rm as}~r\to\infty,
    \label{scat-w-total}
    \ee
where $\Theta$ is the Heaviside step function, i.e.,
    \[\Theta(x):=\left\{\begin{array}{ccc}
    1 & {\rm for} & x\geq 0,\\
    0 & {\rm for} & x<0,\end{array}\right.\]
and $\psi_-(\mathbf{r})$ and $\psi_+(\mathbf{r})$ are respectively the reflected and transmitted waves given by
	\begin{align}
 	&\psi_\pm(\mathbf{r}):=\frac{1}{2\pi}
	\int_{-k}^k dp\, T_\pm(p)\,e^{ipy}e^{\pm i	\omega(p)x}~~{\rm for}~r\to\infty,
	\label{scat-w}	\\
	&T_+(p):=A_+(p)-A_-(p),~~~~~~~~~T_-(p):=B_-(p).
	\label{Tpm}
	\end{align}
The functions $T_\pm(p)$ yield the momentum representation of the reflected and transmitted waves $\psi_\pm(\mathbf{r})$.

For a left-incident plane wave with a definite momentum $k$ along the positive $x$-axis, we have
	\begin{align}
	&A_-(p)=2\pi\delta(p), && B_+(p)=0,
	\label{scat}
	\end{align}
where $\delta$ stands for the Dirac delta-function. Substituting (\ref{scat}) in (\ref{M-def}) and using (\ref{Tpm}), we can express $T_\pm(p)$ in terms of the entries $M_{ij}$ of $\bM$ according to
    \be	
    \begin{aligned}
	&T_+(p)=2\pi(M_{11}-1-M_{12}M_{22}^{-1}M_{21})\delta(p),\\
	&T_-(p)=-2\pi M_{22}^{-1}M_{21}\delta(p).
	\end{aligned}
    \label{Tpm=}
    \ee

Next, we write the scattered wave in its standard form,
	\be
 	\psi_{{\rm scat}}(\mathbf{r})= \sqrt{\frac{i}{kr}}\,e^{ikr} f(\theta)~~~~{\rm as}~r\to\infty,
	\label{scat-w-2}
 	\ee
where $(r,\theta)$ are the polar coordinates of $\mathbf{r}$, and $f(\theta)$ is the scattering amplitude of the potential \cite{adhikari-86}. A careful analysis of the right-hand side of (\ref{scat-w}) gives a closed form asymptotic expression for $\psi_\pm(\mathbf{r})$ (See Appendix~A.) Using this in (\ref{scat-w-total}) and comparing the result with (\ref{scat-w-2}), we obtain
    \be
 	f(\theta)=-\frac{i \omega(p) T_\pm(p)}{\sqrt{2\pi}}=
    -\frac{i k|\cos\theta|}{\sqrt{2\pi}}\,T_\pm(k\sin\theta),				
 	\label{f=}
 	\ee
where $\pm:={\rm sgn}(\cos\theta)$.

Equations (\ref{Tpm=}) and (\ref{f=}) show that $\bM$ includes all the information about the scattering properties of the potential $v$. In other words, the solution of the scattering problem can be achieved by evaluating $\bM$ using (\ref{M=exp}) and computing $T_\pm(p)$ and $f(\theta)$ as given by (\ref{Tpm=}) and (\ref{f=}).

Because the effective Hamiltonian $\bH(x,p)$ scales with the potential $v$, Eq.~(\ref{M=exp}) provides a convenient method of perturbative solution of the scattering problem (similarly to its one-dimensional analog \cite{pra-2014a}). We have checked that if we perform the first order perturbation theory using our approach, we obtain the known expression for the scattering amplitude in the first Born approximation \cite{Lapidus-1982}. Here we do not wish to discuss the utility of our method in perturbation theory. Instead, we demonstrate its application in providing an exact (non-perturbative) solution of the old and difficult problem of delta-function potential in two dimensions. The perturbation series solution of the Lippmann-Schwinger equation \cite{sakurai} for the delta-function potential in two dimensions is plagued with the emergence of infinities. The same problem arises in the study of the corresponding spectral problem and has led to the development of intricate renormalization schemes that amount to the introduction of a length scale for the problem \cite{thorn,exner,henderson-1997,henderson-1998,cabo-1998,phillips,rajeev-99,frederico,Camblong-2001a, camblong-2002}. The problem of divergences also arises in the standard treatment of the delta-function potential in 3D. As we show in the sequel, our approach avoids this problem in both two and three dimensions.

\section{Delta-function potential in 2D}

Consider
    \be
    v(x,y)=\fz\,\delta(x)\delta(y),
    \label{2d-delta}
    \ee
where $\fz$ is a real or complex coupling constant. Then $\bH= \fz\,\omega(p)^{-1}\delta(x)\delta(i\partial_p)\boldsymbol{\cK}/2$ and (\ref{M=exp}) gives
	\begin{align}
  	& \bM=\bI-\frac{i\fz}{2\omega(p)}\,\delta(i\partial_p)\boldsymbol{\cK}.
    \label{2d-delta-M}
  	\end{align}
As we show in Appendix~B, the calculation of $T_\pm(p)$ is more conveniently carried out by substituting (\ref{2d-delta-M}) in (\ref{M-def}) and making use of (\ref{Tpm}) and (\ref{scat}). This gives $T_\pm(p)=-2i\fz/(4+i\fz)\omega(p)$ which together with (\ref{f=}) yield the following exact expression for the scattering amplitude of (\ref{2d-delta}).
	\begin{align}
	&f(\theta)=-\sqrt{\frac{2}{\pi}}\frac{\fz}{4+i\fz}.
	\label{delta2}
	\end{align}
The first Born approximation corresponds to the linear approximation of this equation, i.e., $f(\theta)\approx-\fz/2\sqrt{2\pi}$, which agrees with the results of Ref.~\cite{Lapidus-1982,Camblong-2001a}.

According to (\ref{delta2}), $f(\theta)$ diverges for $\fz=4i$. This marks a zero-width resonance that is associated with what mathematicians call a spectral singularity \cite{prl-2009}. In optical applications \cite{pra-2011a}, $v$ is related to the complex relative permittivity $\varepsilon$ of the medium according to $v=k^2(1-\varepsilon)$. In particular, (\ref{2d-delta}) corresponds to a thin wire of nonmagnetic material that is aligned along the $z$-axis and described by $\varepsilon=1+i\zeta\delta(x)\delta(y)$  for some real parameter $\zeta$. Denoting the unit vector along the positive $z$-axis by $\mathbf{e}_z$, we can easily check that
    \be
    \mathbf{E}(x,y,z,t)=e^{-ickt}\psi(x,y)\mathbf{e}_z
    \label{E=}
    \ee
solves Maxwell's equations for this system provided that $\psi$ satisfies (\ref{sch-eq}) for the delta-function potential (\ref{2d-delta}) with $\fz=-i\zeta k^2$. In terms of $\zeta$, the condition for the emergence of a spectral singularity takes the form $k^2=-4/\zeta$. This shows that the wire begins lasing \cite{pra-2011a} for $k=2/\sqrt{-\zeta}$ whenever $\zeta<0$ (i.e., it is made of gain material), and functions as a coherent perfect absorber \cite{CPA,CPA-Science,longhi-1,longhi-3} for $k=2/\sqrt{\zeta}$ whenever $\zeta>0$ (it is made of lossy material).

\section{A slab laser with a surface line defect}

\begin{figure}
	\begin{center}
	\includegraphics[scale=.5]{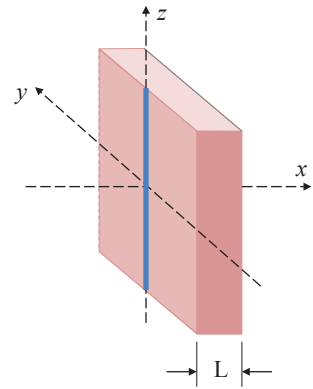}
	\caption{(Color online) A slab of thickness $L$ with a surface line defect shown by the thick blue line.}
	\label{fig2}
	\end{center}
	\end{figure}

As a simple application of the composition property of our transfer matrix, we address the scattering problem for electromagnetic waves interacting with an infinite planar slab of optically active material with a delta-function line defect on one of its faces as depicted in Fig.~\ref{fig2}.

Identifying the location of the slab and the line defect respectively with $0\leq x\leq L$ and $x=y=0$ ($z$-axis) and considering the linearly polarized waves of the form (\ref{E=}), we can reduce the problem to the two-dimensional scattering problem for the potential $v_\delta+\tilde v$, where $v_\delta$ is the delta-function potential (\ref{delta2}) which represents the defect, and $\tilde v$ is a barrier potential of the form
    \be
    \tilde v(x,y):=\left\{\begin{array}{cc}
    \tilde\fz &{\rm for}~0\leq x\leq L,\\
    0 &{\rm otherwise},\end{array}\right.
    \label{v-slab}
    \ee
which represents the slab \cite{pra-2011a}. Here $\tilde\fz:=k^2(1-\varepsilon)$, $\varepsilon$ is the complex relative permittivity of the slab, and $L$ is its thickness.

Because, along the $x$-axis, the support of $v_\delta$ is to the left of that of $\tilde v$, the transfer matrix $\bM$ of $v_\delta+\tilde v$ is given by $\bM=\tilde\bM\bM_\delta$, where $\tilde\bM$ and $\bM_\delta$ are respectively the transfer matrices of $\tilde v$ and $v_\delta$.

We have already computed $\bM_\delta$. It has the form (\ref{2d-delta-M}). We give the details of the calculation of $\tilde\bM$ in Appendix~C. Its entries have the following expression.
    {\small \be
    \begin{aligned}
    &\tilde M_{11}(\omega)=\tilde M_{22}(-\omega)=[\cos(\fn L\omega)+i\fn_+\sin(\fn L\omega)]e^{-i\omega L},\\
    &\tilde M_{12}(\omega)=\tilde M_{21}(-\omega)=i\fn_-\sin(\fn L\omega)e^{-i\omega L},
    \end{aligned}
    \label{M-slab=}
    \ee}%
where $\fn:=\sqrt{1-\tilde\fz/\omega^2}
    =\sqrt{(\varepsilon-1)(k/\omega)^2+1}$ and $\fn_\pm:=(\fn\pm\fn^{-1})/2$. For $p=0$, $\omega=k$, and $\fn$ coincides with the complex refractive index of the slab, namely
$\sqrt\varepsilon$. In this case, (\ref{M-slab=}) gives the known result obtained in one dimension \cite{pra-2011a,pra-2015c}.

Having determined $\bM_\delta$ and $\tilde\bM$, we can use the expression $\bM=\tilde\bM\bM_\delta$ together with (\ref{M-def}), (\ref{Tpm}) and (\ref{scat}) to compute $T_\pm$. As we show in Appendix~D, this gives
    \be
    \begin{aligned}
    \omega\, T_- &=2\pi k[\cX(k)-1]\delta(p)-\frac{i\fz\,\cX(k)\cX(\omega)}{\cY(k)},\\
    \omega\, T_+ &=2\pi k
    \left[\frac{1}{\tilde M_{22}(k)}-1\right]\delta(p)-\frac{i\fz\,\cX(k)}{\cY(k)\tilde M_{22}(\omega)},
    \end{aligned}
    \label{Tm=2d}
    \ee
where
    \begin{align*}
    \cX(\omega)&:=1-\frac{\tilde M_{21}(\omega)}{\tilde M_{22}(\omega)}
    =\frac{2\left(e^{-2i\fn L\omega}+\frac{\fn-1}{\fn+1}\right)}{(\fn+1)\cZ(\omega)},\\
    \cY(k)&:= 2+\frac{i\fz}{\pi}\int_0^k \frac{\cX(\omega)d\omega}{\sqrt{k^2-\omega^2}} ,\\
    \cZ(\omega)&:=e^{-2i\fn L\omega}-\left(\frac{\fn-1}{\fn+1}\right)^2.
    \end{align*}
Substituting (\ref{Tm=2d}) in (\ref{f=}), we find the scattering amplitude for the system.

In the absence of the line defect, $\fz=0$, and the condition for the emergence of a spectral singularity takes the form $\tilde M_{22}(k)=0$ which is equivalent to $\cZ(k)=0$. This equation turns out to be equivalent to the laser threshold and phase conditions for the slab \cite{pra-2011a}. It is easy to see that the inclusion of the line defect yields other means of achieving a spectral singularity. In this case a spectral singularity occurs also for $\tilde M_{22}(\omega)=0$ which is equivalent to $\cZ(\omega)=0$.

Recalling that the gain coefficient $g$ for the slab satisfies $g=-2k\kappa$, where $\kappa$ is the imaginary part of the refractive index $\sqrt\varepsilon$, we can follow the approach of \cite{pra-2011a} to determine the threshold gain from $\cZ(\omega)=0$. Using the fact that for generic active material, $|\kappa|\ll \eta-1$, where $\eta$ is the real part of $\sqrt\varepsilon$, we then find the following expression for the threshold gain coefficient
    \be
    g=\frac{4\sqrt{\eta^2-\sin^2\theta}}{\eta L}
    \ln\left[\frac{\sqrt{\eta^2-\sin^2\theta}+|\cos\theta|}{
    \sqrt{\eta^2-1}}\right].
    \label{g=}
    \ee
This is precisely the laser threshold condition for the oblique TE modes of the slab in the absence of the defect, if we identify $\theta$ with the incidence angle \cite{pra-2015a}. Notice however that here we have a normally incident TE wave and $\theta$ corresponds to the direction of the scattered wave.

\begin{figure}
	\begin{center}
	\includegraphics[scale=.65]{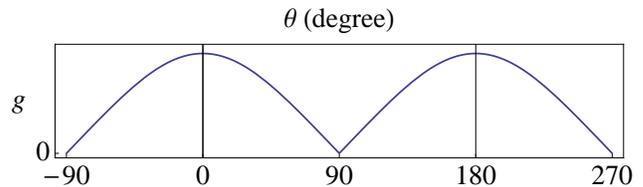}
	\caption{Graph of the threshold gain coefficient  $g$ as a function of $\theta$ for a generic slab laser with a surface line defect.}
	\label{fig3}
	\end{center}
	\end{figure}

Fig.~\ref{fig3} shows a plot of the threshold gain coefficient (\ref{g=}) as a function of $\theta$ for a generic value of $\eta>1$.
As seen from this graph, $g$ attains its maximum value at $\theta=0,180^\circ$. If $\theta$ deviates from these values, $g$ decreases monotonically until it reaches its minimum value, which is zero, at $\theta=\pm 90^\circ$. This shows that if the slab consists of a gain medium, irrespective of how small its gain coefficient is, the defect will make it radiate laser light with a wavevector in some direction. This is actually easy to justify. The presence of the defect makes the waves scatter in different directions. Those that are scattered through angles closer to $\theta=\pm 90^\circ$ follow a longer optical path within the slab and lase more easily.

\section{Transfer matrix in 3D}

First we introduce some useful conventions and notions. We take the scattering axis of our system to be aligned along the $z$-axis of a cartesian coordinate system, label the position vector $\mathbf{r}$ with its cartesian coordinates $(x,y,z)$, and set $\vec{\rho}:=(x,y)$  and $(\vec{\rho},z):=(x,y,z)$. In general overarrows identify two-dimensional vectors lying in the $x$-$y$ plane.

Now, consider a potential $v:\R^3\to\C$ such that for $z\to\pm\infty$ every solution of the Schr\"odinger equation,
    \be
    \left[-\partial_x^2-\partial_y^2-\partial_z^2+v(x,y,z)\right]\psi(x,y,z)=k^2\psi(x,y,z),
    \label{sch-3D-z}
    \ee
has the asymptotic form
    \bea
    \frac{1}{(2\pi)^2}
	\int_{\sD_k} d^2p\, e^{i\vec p\cdot\vec\rho}\left[A_\pm(\vec p)e^{i\omega(\vec p)z}+
	B_\pm(\vec p) e^{-i\omega(\vec p)z}\right],\nn
	\eea
where $\sD_k$ stands for the disc $\big\{\vec p\in\R^2\big||\vec p|\leq k\big\}$, $A_\pm$ and $B_\pm$ are coefficient functions vanishing outside $\sD_k$, and $\omega(\vec p):=\sqrt{k^2-\vec{p}^{\:2}}$.
We can use (\ref{M-def}) to define a transfer matrix $\bM$ for the potential $v$, and check that it satisfies (\ref{M=exp}) if we replace $(x,p)$ by $(z,\vec p)$, and define $\bH(z,\vec p)$ according to
    \be
	\boldsymbol{\bH}(z,\vec p):=\frac{1}{2\omega(\vec p)} e^{-i\omega(\vec p)z\boldsymbol{\sigma}_3}
	v(i\vec\partial_p,z)\,\boldsymbol{\cK}\,e^{i\omega(\vec p)z\boldsymbol{\sigma}_3}.
	\label{sH=3D}
 	\ee
Here $\vec\partial_p:=(\partial_{p_x},\partial_{p_y})$, $(p_x,p_y)$ are cartesian coordinates of $\vec p$, and $v(i\vec\partial_p,z):=v(i\partial_{p_x},i\partial_{p_y},z)$. For details of the derivation of (\ref{sH=3D}), see Appendix~E.

According to (\ref{sH=3D}), $\boldsymbol{\bH}(z,\vec p)$ vanishes for the values of $z$ for which $v(x,y,z)=0$. This together with Eq.~(\ref{M=exp}) ensure that $\bM$ satisfies the composition property (\ref{comp-rule}).

In order to relate $\bM$ to the scattering amplitude for the potential $v$, we introduce $T_\pm(\vec p)$ by changing $p$ to $\vec p$ in (\ref{Tpm}). These have the same properties as their two-dimensional analog. We can express them in terms of $\bM$ by enforcing (\ref{M-def}) and
setting
    \begin{align*}
    &A_-(\vp)=4\pi^2\delta(p_x)\delta(p_y),
    &&B_+(\vp)=0.
    \end{align*}
In terms of $T_\pm(\vec p)$  the reflected ($\psi_-$) and transmitted ($\psi_+$) waves take the form
    \be
    \psi_\pm(\mathbf{r})=\frac{1}{4\pi^2}\int_{\sD_k}d^2p\, T_\pm(\vec p)
    e^{i\vec p\cdot\vec\rho}e^{\pm i\omega(\vec p)z}
    ~~~~{\rm as}~r\to\infty.
    \label{scat-w-3D}
    \ee
These in turn give the scattered wave $\psi_{\rm scat}$ according to (\ref{scat-w-total}) provided that we replace $x$ by $z$.

Next, we recall that $\psi_{\rm scat}(\mathbf{r})=f(\vartheta,\varphi)\,e^{ikr}/r$, where $f$ is
the scattering amplitude and $(r,\vartheta,\varphi)$ are the spherical coordinates of $\mathbf{r}$ with $\vartheta$ and $\varphi$ respectively denoting the polar and azimuthal angles. Comparing this expression with the one given by (\ref{scat-w-total}) and (\ref{scat-w-3D}) and using an argument given in Appendix~F, we find
    \be
    f(\vartheta,\varphi)=-\frac{ik|\cos\vartheta|}{2\pi}\,T_\pm(k\sin\vartheta\cos\varphi,k\sin\vartheta\sin\varphi),
    \label{f=3D}
    \ee
where $\pm:={\rm sgn}(\cos\vartheta)$.

\section{Delta-function potential in 3D}

For the delta-function potential
    \be
    v(x,y,z)=\fz\,\delta(x)\delta(y)\delta(z),
    \label{delta-3D-1}
    \ee
Eq.~(\ref{sH=3D}) gives $\boldsymbol{\bH}(z,\vp)=\fz\,\omega(\vec p)^{-1}\delta(z)\delta(i\partial_{p_x})\delta(i\partial_{p_y})\boldsymbol{\cK}/2$. Substituting this relation in (\ref{M=exp}) and making use of (\ref{M-def}), (\ref{scat}), and (\ref{f=3D}), we find
    \begin{align}
    &\bM=\bI-\frac{i\fz}{2\omega(\vp)}\,\boldsymbol{\cK}\delta(i\partial_{p_x})\delta(i\partial_{p_y}),
    \label{M=delta-3D}\\
    &f(\vartheta,\varphi)=\frac{-\fz}{4\pi +ik\fz}.
    \label{f=delta-3D}
    \end{align}

In view of the latter relation the scattering length, i.e., $\xi:=-\lim_{k\to 0}f$, is given by $\xi=\fz/4\pi$. Therefore, we can express (\ref{f=delta-3D}) as $f=-(\xi^{-1}+ik)^{-1}$. This coincides with Eq.~(9) of Ref.~\cite{rajeev-99}, which is obtained for general unitary interactions under the assumption that the range of the interaction is much smaller than the wavelength of the incident wave.

The standard physical application of delta-function potential is in the study of point defects in condensed matter systems. For a charged particle moving in a conducting medium, the presence of a point defect yields a non-vanishing resistivity, which is related to the scattering cross section of the defect $|f(\vartheta,\varphi)|^2$. Eq.~(\ref{f=delta-3D}) suggests that it is an isotropic function of space that is inversely proportional to $k^2+\mu^2$ where $\mu:=4\pi/|\fz|$.

\section{Concluding Remarks}

Scattering of waves in one dimension is most conveniently studied in the framework of the transfer matrix formulation of scattering theory. This is mainly because of the practical advantages stemming from the composition property of the transfer matrix. Despite the enormous interest in the scattering of waves in two- and three-dimensions, there has been no successful attempt at developing a general transfer matrix formulation of scattering theory in these dimensions. In this article, we achieve this goal by introducing a genuine multi-dimensional notion of transfer matrix which enjoys the niceties of its one-dimensional analog.

In order to elucidate the practical aspects of our approach, we have explored its application in addressing the scattering problem for delta-function potentials in two and three dimensions. It is known that a proper treatment of the spectral problem for delta-function potential in two dimensions requires a coupling constant renormalization which makes it run with $k$ and introduces a length scale for the problem \cite{rajeev-99,camblong-2002}. As far as its physical applications are concerned, the delta-function potential is an abstraction for a physical potential whose range is much smaller than the relevant wavelengths. This provides an explanation for the length scale arising from the renormalization procedure and applies for the delta-function potential in three dimensions. It is remarkable that the direct application of our approach to delta function potentials yield finite results that do not rely on any approximation or renormalization schemes.

The composition property of the transfer matrix allows for extending our results to potentials consisting of finitely many delta-functions. In two dimensions, these describe finite arrays of parallel thin wires whose scattering properties are of direct interest in the study of metamaterials \cite{meta1,meta2} and nanophotonic devices \cite{nano1,nano2,nano3}.  In three dimensions, delta-function potentials model point-like defects. This marks the significance of our approach for dealing with scattering problems in condensed matter physics. It is also not difficult to envisage a generalization of this approach that is capable of treating scattering of waves inside a waveguide. This will be of importance in the study of transport phenomena in mesoscopic systems \cite{MK}.

To demonstrate the effectiveness of our approach in dealing with a realist physical system, we have used it to describe the scattering of electromagnetic ways from an optically active slab that has a surface line defect. Our results allow for the determination of the laser threshold condition for such a slab and reveal the fact that for any positive value of the gain coefficient the presence of the defect makes the slab begin emitting laser light in some direction. It is remarkable that this phenomenon does not depend on the strength of the delta-function interaction modeling the defect.

The main impact of the multidimensional transfer matrix formulation of scattering theory that we outline in this article will be in its numerical implementations. This is not only because of the composition property of the transfer matrix, which allows for the reduction of the scattering problem for a given potential to those of its truncations, but also because any discretization of this approach would yield matrix representations for the entries of the transfer matrix operator that could be substituted in (\ref{Tpm=}) to determine $T_\pm$ and consequently the scattering amplitude.

\subsection*{Acknowledgments}

This project was initiated during F.~L.'s visit to Ko\c{c} University Aug.\ 11-24, 2015. We are indebted to the Turkish Academy of Sciences (T\"UBA) for providing the financial support which made this visit possible and to Teoman Turgut for helpful discussions. This work has been supported by the Scientific and Technological Research Council of Turkey (T\"UB\.{I}TAK) in the framework of the project no: 112T951 and by T\"UBA.

\section*{Appendices}

In the following we give the details of various calculations that we refer to in the text of this article.

\subsection*{A: Expressing $\psi_\pm(\mathbf{r})$ and $f(\theta)$ in terms of $T_\pm$ in 2D}

We wish to determine the asymptotic behavior of the factor $e^{ipy}e^{\pm i\omega(p)x}$ appearing in (\ref{scat-w}) for $r\to\infty$. First, we use $(x,y)=(r\cos\theta, r\sin\theta)$ and the identity
	\be
	\cos(\alpha\pm\beta)=\cos\alpha\cos\beta\mp\sin\alpha\sin\beta,
	\label{cos-iden}
	\ee
to express this quantity in the form
        \be
        e^{ipy}e^{\pm i\omega(p)x}=e^{ikr\cos(\theta_p-\theta^\pm)},
        \label{zq1}
        \ee
where $\theta^+:=\theta$, $\theta^-:=\pi-\theta$, and $\theta_p$ is the real variable determined by the conditions
        \begin{align}
        &|\theta_p|\leq\frac{\pi}{2},
        &&p=k\sin\theta_p,
        &&\omega(p)=k\cos\theta_p.
        \end{align}

    Next, we recall from Ref.~\cite{Lapidus-1982} that, for $r\to\infty$,
        \be
        e^{ikr\cos\phi}\to\sqrt{\frac{2}{\pi k r}}\sum_{m=0}^\infty\epsilon_m i^m \cos\left(kr-\frac{m\pi}{2}-\frac{\pi}{4}\right)  \cos(m \phi),
        \label{zq2}
        \ee
     where $\epsilon_0=1$ and $\epsilon_m=2$ for $m\geq 1$. If we use the identity (\ref{cos-iden}) to expand the factor
     $\cos\left(kr-\frac{m\pi}{2}-\frac{\pi}{4}\right)$ in (\ref{zq2}),
     split the series into the sum of the terms with odd and even values of $m$, and notice that for every integer $\ell$, $\sin(\ell\pi)=\cos(\ell\pi+\frac{\pi}{2})=0$ and $\sin(\ell\pi+\frac{\pi}{2})=\cos(\ell\pi)=(-1)^\ell$, we have
        \begin{align}
        &\sum_{m=0}^\infty\epsilon_m i^m \cos\left(\xi-\frac{m\pi}{2}\right)\cos(m \phi)\nn\\
        &=\cos\xi\sum_{\ell=0}^\infty\epsilon_{2\ell}
        \cos(2\ell\phi)+i\sin\xi\sum_{\ell=0}^\infty\epsilon_{2\ell+1}
        \cos[(2\ell+1)\phi]\nn\\
        &=\frac{e^{i\xi}}{2}\sum_{m=0}^\infty\epsilon_m\cos(m\phi)+
        \frac{e^{-i\xi}}{2}\sum_{m=0}^\infty(-1)^m\epsilon_m\cos(m\phi)\nn\\
        &=\pi\left[e^{i\xi}\delta(\phi)+e^{-i\xi}\delta(\phi+\pi)\right],
        \label{zq3}
        \end{align}
     where we have introduced $\xi:=kr-\frac{\pi}{4}$ and employed the identities:
        \begin{align*}
        &\sum_{m=0}^\infty \epsilon_m\cos(m\phi)=\sum_{n=-\infty}^\infty e^{i n\phi}=2\pi\delta(\phi),\\
        &\sum_{m=0}^\infty (-1)^m\epsilon_m\cos(m\phi)=2\pi\delta(\phi+\pi).
        \end{align*}

     Next, we change the variable of integration in (\ref{scat-w}) to $\theta_p$ and use (\ref{zq1}) -- (\ref{zq3}) to evaluate this integral. {Note that for both $\theta\in[-\frac{\pi}{2},\frac{\pi}{2}]$ and $\theta\in[\frac{\pi}{2},\frac{3\pi}{2}]$, we have $\theta^\pm\in[-\frac{\pi}{2},\frac{\pi}{2}]$.} Therefore in both cases it is the first delta-function in Eq.\eqref{zq3} that contributes to the integral. The result is
     	\be
	\psi_\pm(\mathbf{r})=\frac{k\cos\theta^\pm e^{-i\pi/4}e^{ikr}}{\sqrt{2\pi k r}}T_\pm(k\sin\theta^\pm).
	\ee
Substituting this relation in (\ref{scat-w-total}) and comparing the result with (\ref{scat-w-2}), we obtain
	\[ f(\theta)=\frac{-ik\cos(\theta^\pm)}{\sqrt{2\pi}}T_\pm(k\sin\theta^\pm)~~~{\rm for}~
	\pm\cos\theta>0.\]
It is not difficult to see that we can express this relation as Eq.~(\ref{f=}).

\subsection*{B: Calculation of $T_\pm$ for delta-function potential in 2D}

For $v(x,y)=\fz\delta(x)\delta(y)$,  the effective Hamiltonian takes the form $\bH(x,p)=\fz\,\omega(p)^{-1}\delta(x)\delta(i\partial_p)\,\boldsymbol{\cK}/2$. Substituting this equation in (\ref{M=exp}) and noting that $\boldsymbol{\cK}^2=\mathbf{0}$, we obtain Eq.~(\ref{2d-delta-M}), which reads
$\bM=1-i\fz\omega(p)^{-1}\delta(i\partial_p)\boldsymbol{\cK}/2$.

Next, we recall that
    \begin{align}
    &\boldsymbol{\cK}=\boldsymbol{\sigma}_3+i\boldsymbol{\sigma}_2=\left[\begin{array}{cc}
    1 & 1 \\ -1 & -1\end{array}\right],
    \label{zq4}\\
    &\delta(i\partial_p) \phi(p)=\frac{1}{2\pi}\int_{-k}^k dp\, \phi(p)=:\tilde \phi,
    \label{zq5}
    \end{align}
where $\phi:\R\to\C$ is a test function such that $\phi(p)=0$ for $|p|>k$. If we insert (\ref{2d-delta-M}) and (\ref{scat}) in (\ref{M-def}) and make use of (\ref{zq4}) and (\ref{zq5}), we find
    \be
    \left[\begin{array}{c}A_+(p)\\ 0\end{array}\right]=\left[\begin{array}{c}A_-(p)\\ B_-(p)\end{array}\right]-\frac{i\fz}{2\omega(p)}(1+\tilde B_-)\left[\begin{array}{c} 1\\ -1\end{array}\right]
    \nn
    \ee
In light of this equation and (\ref{Tpm}),
     \be
     T_\pm(p)=-\frac{i\fz(\tilde B_-+1)}{2\omega(p)}.
     \label{2d-single-delta}
     \ee
Because $T_-(p)=B_-(p)$, this relation in particular implies $B_-(p)=-i\fz (\tilde B_-+1)/2\omega(p)$. Applying $\delta(i\partial_p)$ to both sides of this equation and employing Eq.~(\ref{zq5}) and
    \be
    \delta(i\partial_p)\omega^{-1}(p)=\frac{1}{2\pi}\int_{-k}^k\frac{dp}{\sqrt{k^2-p^2}}
    =\frac{1}{2},
    \label{id-1-2}
    \ee
we obtain $\tilde B_-=-i\fz/(4+i\fz)$. Inserting this in \eqref{2d-single-delta} gives (\ref{delta2}).

\subsection*{C: Calculation of the transfer matrix for a slab without a defect}

Consider a scattering potential $v$ in two-dimensions that does not depend on $y$. Then Eq.~(\ref{sch-eq-FT}) and the effective Hamiltonian take the form
        \begin{align*}
        & \left[-\partial_x^2+v(x)\right]\tilde\psi(x,p)=\omega^2(p)\tilde\psi(x,p),\\
        &\bH(x,p)=\frac{1}{2\omega(p)} v(x)\,e^{-i\omega(p)x\boldsymbol{\sigma}_3}
    	\boldsymbol{\cK}\,e^{i\omega(p)x\boldsymbol{\sigma}_3},
        \end{align*}
    respectively. For the potential given by (\ref{v-slab}), these reduce to their one-dimensional analogs for a rectangular barrier potential \cite{pra-2015c} provided that we replace $k$ with $\omega(p)$ in the latter. In particular, making this change in Eq.~(32) of Ref.~\cite{pra-2015c}, we obtain Eq.~(\ref{M-slab=}). Alternatively, one can perform the $x$-dependent transformation of two-component state vectors: $\Psi(x)\to \check\Psi(x):= e^{i\omega(p)x\boldsymbol{\sigma}_3}\Psi(x)$ that gives rise to an $x$-independent effective Hamiltonian $\check{\bH}(p)$, obtain the corresponding evolution operator, and transform back to compute $\bM$. This also leads to Eq.~(\ref{M-slab=}).

\subsection*{D: Calculation of $T_\pm$ for a slab with a line defect}

As we noted in the text of the paper, by virtue of the composition property (\ref{M-def}), the transfer matrix for the potential $v_\delta+\tilde v$ has the form
        \be
        \bM=\tilde\bM\,\bM_{\delta},
        \label{M=line}
        \ee
    where $\bM_\delta$ and $\tilde\bM$ are respectively given by (\ref{2d-delta-M}) and (\ref{M-slab=}). Substituting (\ref{2d-delta-M}) in (\ref{M=line}) and enforcing (\ref{M-def}) and (\ref{scat}), we find
        \bea
        \label{A+barrier-3D}
        A_+(p)&=&\tilde M_{11}(\omega)\left[2\pi\delta(p)-\frac{i\fz}{2\omega}(\tilde B_-+1)\right]+\nn\\
        &&\tilde M_{12}(\omega)\left[B_-(p)+\frac{i\fz}{2\omega}(\tilde B_-+1)\right],\\
        B_-(p)&=&-\tilde M_{22}^{-1}(k)\,\tilde M_{21}(k)2\pi\delta(p)+
        \label{B+barrier-3D}\\
        &&\frac{i\fz}{2\omega(p)}(\tilde B_-+1)\left(\tilde M_{22}^{-1}(\omega)\,\tilde M_{21}(\omega)-1\right),\nn
        \eea
    where we have employed (\ref{zq5}).  Next, we determine $\tilde B_-$ by operating $\delta(i\partial_p)$ on both sides of \eqref{B+barrier-3D} and making use of (\ref{id-1-2}) and (\ref{M-slab=}). This gives $\tilde B_-+1= 2\cX(k)/\cY(k)$ where $\cX(k)$ and $\cY(k)$ are defined in the text of the paper. Substituting this expression for $\tilde B_-+1$ in (\ref{A+barrier-3D}) and (\ref{B+barrier-3D}) and using (\ref{Tpm}) and (\ref{M-slab=}) we obtain (\ref{Tm=2d}).

\subsection*{E: Expressing $\bM$ in terms of an evolution operator in 3D}

Following the approach we pursued for treating the two-dimensional case, we first take the Fourier transform of both sides of the Schr\"odinger equation (\ref{sch-3D-z}) over $x$ and $y$. This gives
        \be
        -\partial_z^2\tilde\psi(\vp,z)+v(i\vd_p,z)\tilde\psi(\vp,z)=\omega(\vp)^2\tilde\psi(\vp,z),
        \label{caution-higher-d}
        \ee
where
    \bea
    \tilde\psi(\vp,z)&:=&\int_{-\infty}^\infty dx \int_{-\infty}^\infty dy\, e^{-i(xp_x+yp_y)}\psi(x,y,z)\nn
    \eea
is the Fourier transform of $\psi(x,y,z)$ over $x$ and $y$. Next, we introduce the two-component state vector
        \begin{align*}
        \Psi(z,\vec p)&:=\frac{1}{2}\left[\begin{array}{c}
        e^{-i\omega(\vec p)z}\Big(\tilde\psi(\vec p,z)-\displaystyle\frac{i\partial_z \tilde\psi(x,\vec p)}{\omega(\vec p)}\Big)\\[6pt]
        e^{i\omega(\vec p)z}\Big(\tilde\psi(\vec p,z)+\displaystyle\frac{i\partial_z \tilde\psi(\vec p,z)}{\omega(\vec p)}\Big)\end{array}\right],
        \end{align*}
and check that it has the same properties as its analogue in two-dimensions. In particular, (\ref{caution-higher-d}) is equivalent to $i\partial_z\Psi(z,\vec p)=\bH(z,\vec p)\Psi(z,\vec p)$ for the effective Hamiltonian $\bH(z,\vec p)$ given by (\ref{sH=3D}), and
$\Psi(z,\vec p)\to\left[\begin{array}{c} A_\pm(\vec p)\\ B_\pm(\vec p)\end{array}\right]$  for $z\to\pm\infty$. These together with (\ref{M-def}) imply $\bM=\bU(\infty,-\infty)$, where $\bU(z,z_0)$ is the evolution operator for the effective Hamiltonian $\bH(z,\vec p)$, i.e., $\bU(z,z_0):=\sT\exp\int_{z_0}^z -i \bH(z',\vec p) dz'$.

\subsection*{F: Expressing $\psi_\pm(\mathbf{r})$ and $f(\theta)$ in terms of $T_\pm$ 3D}

We wish to determine the asymptotic form of the factor $e^{i\vec p\cdot\vec\rho}e^{\pm i\omega(\vec p)z}$ that appears in Eq.~(\ref{scat-w-3D}). First, we introduce $ \theta^+:=\theta$,  $\theta^-:=\pi-\theta$, and
${\bf r}_\pm:=r(\cos\theta^\pm,\sin\theta^\pm\cos\phi,\sin\theta^\pm\sin\phi)$, where $(r,\theta,\phi)$ are the standard spherical coordinates, i.e., $z=r\cos\theta$ and $\vec \rho=r\sin\theta(\cos\phi,\sin\phi)$. Then, in view of ${\bf p}:=(\omega(p),\vec p)$, it is not difficult to see that
        \be
        e^{i\vec p\cdot\vec\rho}e^{\pm i\omega(\vec p)z}=e^{i{\bf p}\cdot{\bf r}_\pm}.
        \label{zq5.5}
        \ee
We can use the spherical wave expansion of plane waves \cite{Jackson} to express the right-hand side of this equation in the form
        \be
        e^{i{\bf p}\cdot{\bf r}_\pm}=4\pi\sum_{\ell=0}^\infty i^\ell j_\ell(k r)\sum_{m=-\ell}^\ell Y_{\ell m}^*(\theta^\pm,\phi)Y_{\ell m}(\theta_p,\phi_p),
        \label{zq6}
        \ee
where $j_\ell$ and $Y_{\ell m}$ are respectively the spherical Bessel functions and spherical harmonics, and the parameters $\theta_p\in\left[0,\frac{\pi}{2}\right]$ and $\phi_p\in[0,2\pi)$ are determined by
     \begin{align}
     \omega(\vp)=k\cos\theta_p, &&\vp=k\sin\theta_p\left(\cos\phi_p,\sin\phi_p\right).
     \end{align}
 Next, we recall the following asymptotic expression for the spherical Bessel functions and the completeness relation for the spherical harmonics \cite{Jackson}.
     \be
     j_\ell(x)\to\frac{\sin\left(x-\frac{\ell\pi}{2}\right)}{x}=\frac{(-i)^\ell e^{ix}-i^\ell e^{-ix}}{2ix}~~{\rm for}~~x\gg\ell,
     \label{zq7}
     \ee
     \be
     \sum_{\ell=0}^\infty\sum_{m=-\ell}^\ell Y_{\ell m}^*(\theta,\phi)
     Y_{\ell m}(\theta_p,\phi_p)=\delta(\phi_p-\phi)\delta(\cos \theta_p-\cos\theta).\nn
     \ee
Because $Y_{\ell m}(\pi-\theta,\pi+\phi)=(-1)^\ell Y_{\ell m}(\theta,\phi)$, the latter relation implies
     \begin{align}
     &\sum_{\ell=0}^\infty\sum_{m=-\ell}^\ell (-1)^\ell Y_{\ell m}^*(\theta,\phi)Y_{\ell m}(\theta_p,\phi_p)\nn\\
     &\hspace{2cm}=\delta(\pi+\phi_p-\phi)\delta(\cos \theta_p+\cos\theta).~~~~~~
     \label{zq8}
     \end{align}
 In light of (\ref{zq5.5})--(\ref{zq8}), we have the asymptotic formula
     \begin{align}
     &e^{i\vec p\cdot\vec\rho}e^{\pm i\omega(\vec p)z}
     \to -\frac{2i\pi}{kr}\Big[e^{ikr}\delta(\phi_p-\phi)\delta(\cos\theta_p-\cos
     \theta^\pm) \nn\\
     & -e^{-ikr}\delta(\phi_p-\phi+\pi)\delta(\cos\theta_p+\cos \theta^\pm)\Big]~~{\rm for}~r\to\infty,
     \label{3Dasym}
     \end{align}
 which together with $d^2p=k^2\sin\theta_p\cos\theta_p d\theta_p d\phi_p$ allow us to evaluate the integral in (\ref{scat-w-3D}). Because for both $\theta \in[0,\frac{\pi}{2}]$ and $\theta \in[\frac{\pi}{2},\pi]$, we have $\theta^\pm\in[0,\frac{\pi}{2}]$, in both cases it is the first term on the right-hand side of \eqref{3Dasym} that contributes to this integral. The result is $\psi_\pm(\mathbf{r})=f(\theta,\phi) e^{ikr}/r$, where
    \be
    f(\theta,\phi):=-\frac{i}{2\pi}\omega(\vec k)T_\pm(\vec k),
    \label{3d-f}
    \ee
 $\pm:={\rm sgn}(\cos\theta)$, $\omega(\vec k)=k\left|\cos\theta\right|$, and $ \vec k=k\sin\theta\left(\cos\phi,\sin\phi\right)$. In view of (\ref{scat-w-total}),  (\ref{3d-f}) is equivalent to  (\ref{f=3D}).

\ed